# A Survey of Biometric keystroke Dynamics: Approaches, Security and Challenges

Mrs. D. Shanmugapriya
Dept. of Information Technology
Avinashilingam University for Women
Coimbatore, Tamilnadu, India
ds_priyaa@rediffmail.com

Dr. G. Padmavathi
Dept. of Computer Science
Avinashilingam University for Women
Coimbatore, Tamilnadu, India
ganapathi.padmavathi@gmail.com

*Abstract*— **Biometrics technologies are gaining popularity today since they provide more reliable and efficient means of authentication and verification. Keystroke Dynamics is one of the famous biometric technologies, which will try to identify the authenticity of a user when the user is working via a keyboard. The authentication process is done by observing the change in the typing pattern of the user. A comprehensive survey of the existing keystroke dynamics methods, metrics, different approaches are given in this study. This paper also discusses about the various security issues and challenges faced by keystroke dynamics.**

## I. INTRODUCTION

The first and foremost step in preventing unauthorized access is user Authentication. User authentication is the process of verifying claimed identity. The authentication is accomplished by matching some short-form indicator of identity, such as a shared secret that has been pre-arranged during enrollment or registration for authorized users. This is done for the purpose of performing trusted communications between parties for computing applications.

Conventionally, user authentication is categorized into three classes [17]:

- Knowledge - based,
- Object or Token - based,
- Biometric - based.

The following Figure 1. shows the different classification of user authentication methods.

The knowledge-based authentication is based on something one knows and is characterized by secrecy. The examples of knowledge-based authenticators are commonly known passwords and PIN codes. The object-based authentication relies on something one has and is characterized by possession.

Behavioural characteristics are related to what a person does, or how the person uses the body. Voiceprint, gait

Traditional keys to the doors can be assigned to the object-based category. Usually the token-based approach is combined with the knowledge-based approach. An example of this combination is a bankcard with PIN code. In knowledge-based and object-based approaches, passwords and tokens can be forgotten, lost or stolen. There are also usability limitations associated with them. For instance, managing multiple passwords / PINs, and memorizing and recalling strong passwords are not easy tasks. Biometric-based person recognition overcomes the above mentioned difficulties of knowledge-based and object based approaches.

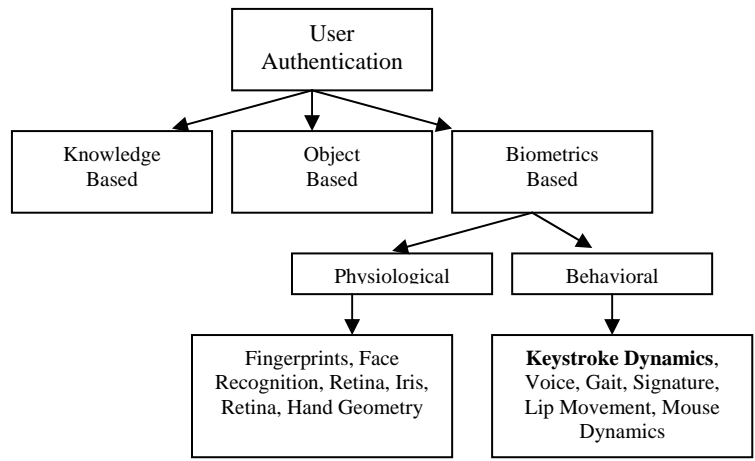

Figure 1.  Classification of User Authentication approaches

Biometric technologies are defined as automated methods of verifying or recognizing the identity of a living person based on a physiological or behavioral characteristics [2]. Biometrics technologies are gaining popularity due to the reason that when used in conjunction with traditional methods of authentication they provide an extra level of security. Biometrics involves something a person is or does. These types of characteristics can be approximately divided into physiological and behavioural types [17]. Physiological characteristics refer to what the person is, or, in other words, they measure physical parameters of a certain part of the body. Some examples are Fingerprints, Hand Geometry, Vein Checking, Iris Scanning, Retinal Scanning, Facial Recognition, and Facial Thermogram. recognition, Signature Recognition, Mouse Dynamics and keystroke dynamics, are good examples of this group.

Keystroke dynamics is considered as a strong behavioral Biometric based Authentication system [1]. It is a process of analyzing the way a user types at a terminal by monitoring the keyboard in order to identify the users based on habitual typing rhythm patterns. Moreover, unlike other biometric systems, which may be expensive to implement, keystroke dynamics is almost free as the only hardware required is the keyboard.

This paper surveys various keystroke dynamics approaches and discusses about the security provided by keystroke dynamics. The paper is structured as follows: the next section gives the identification and verification in keystroke dynamics. Section III explains the methods and metrics of keystroke dynamics. Section IV discusses the various performance measures. Existing approaches are discussed in Section V. The Sixth and Seventh Sections discuss about the security and challenges of keystroke dynamics respectively and final section concludes the topic.

## II. IDENTIFICATION AND VERIFICATION

Keystroke dynamics systems can run in two different modes [2] namely the Identification mode or Verification mode. Identification is the process of trying to find out a person's identity by examining a biometric pattern calculated from the person's biometric features. A larger amount of keystroke dynamics data is collected, and the user of the computer is identified based on previously collected information of keystroke dynamics profiles of all users. For each of the users, a biometric template is calculated in this training stage. A pattern that is going to be identified is matched against every known template, yielding either a score or a distance describing the similarity between the pattern and the template. The system assigns the pattern to the person with the most similar biometric template. To prevent impostor patterns (in this case all patterns of persons not known by the system) from being correctly identified, the similarity has to exceed a certain level. If this level is not reached, the pattern is rejected. Identification with keystroke dynamics means that the user has to be identified without additional information besides measuring his keystroke dynamics.

A person's identity is checked in the verification case. The pattern that is verified is only compared with the person's individual template. Keystroke verification techniques can be classified as either static and dynamic or continuous [22]. Static verification approaches analyze keystroke verification characteristics only at specific times providing additional security than the traditional username/password. For example, during the user login sequence. Static approaches provide more robust user verification than simple passwords but the detection of a user change after the login authentication is impossible. Continuous verification, on contrary, monitors the user's typing behavior throughout the course of the interaction. In the continuous process, the user is monitored on a regular basis throughout the time he/she is typing on the keyboard, allowing a real time analysis [21]. It means that even after a successful login, the typing patterns of a person are constantly analyzed and when they do not match the user's profile, access to the system is blocked.

## III. METHODS AND METRICS FOR KEYSTROKE DYNAMICS

Previous studies [3, 5, 7, 10, 15] have identified a selection of data acquisition techniques and typing metrics upon which keystroke analysis can be based. The following section summarizes the basic methods and metrics that can be used.

*Static at login* – Static keystroke analysis authenticates a typing pattern based on a known keyword, phrase or some other predetermined text. The typing pattern captured is compared against a previously recorded typing patterns stored during system enrollment.

*Periodic dynamic* – Dynamic keystroke analysis authenticates a user on the basis of their typing during a logged session. The data, which is captured in the logged session, is then compared to an archived typing pattern to determine the deviations. In a periodic configuration, the authentication can be constant; either as part of a timed supervision.

*Continuous dynamic* – Continuous keystroke analysis extends the data capturing to the entire duration of the logged session. The continuous nature of the user monitoring offers significantly more data upon which the authentication judgment is based. Furthermore, an impostor may be detected earlier in the session than under a periodically monitored implementation.

*Keyword-specific* – Keyword-specific keystroke analysis extends the continuous or periodic monitoring to consider the metrics related to specific keywords. Extra monitoring is done to detect potential misuse of sensitive commands. Static analysis could be applied to specific keywords to obtain a higher confidence judgment.

*Application-specific* – Application-specific keystroke analysis further extends the continuous or periodic monitoring. It may be possible to develop separate keystroke patterns for different applications.

In addition to a range of implementation scenarios, there are also a variety of possible keystroke metrics. The Following are the metrics widely used by keystroke dynamics.

*Digraph latency* – Digraph latency is the metric that is most commonly used and it typically measures the delay between the key-up and the subsequent key-down events, which are produced during normal typing (e.g. pressing letter T-H).

*Trigraph latency* – Trigraph latency extends the digraph latency metric to consider the timing for three successive keystrokes (e.g. pressing letter T-H-E).

*Keyword latency* – Keyword latencies consider the overall latency for a complete word or may consider the unique combinations of digraph / trigraphs in a word-specific context.

## IV. PERFORMANCE MEASURES

Performance of Keystroke analysis is typically measured in terms of various error rates [13], namely False Accept Rate (FAR) and False Reject Rate (FRR). FAR is the probability of an impostor posing as a valid user being able to successfully gain access to a secured system. In statistics, this is referred to as a Type II error. FRR measures the percent of valid users who are Keystroke Dynamics-based Authentication rejected as impostors. In statistics, this is referred to as a Type I error. Both error rates should ideally be 0%. From a security point of view, type II errors should be minimized that is no chance for an unauthorized user to login. However, type I errors should also be infrequent because valid users get annoyed if the system rejects them incorrectly. One of the most common measures of biometric systems is the rate at which both accept and reject errors are equal. This is known as the Equal Error Rate (EER), or the Cross-Over Error Rate (CER). The value indicates that the proportion of false acceptances is equal to the proportion of false rejections. The lower the equal error rate value, the higher the accuracy of the biometric systems.

## V. KEYSTROKE ANALYSIS APPROACHES

A number of studies [5,7,10,12,20-22,27,28] have been performed in the area of keystroke analysis since its conception. There are two main keystroke analysis approaches for the purposes of identity verification. They are statistical techniques and neural networks techniques. Some are the combinations of both the approaches. The basic idea of the statistical approach is to compare a reference set of typing characteristics of a certain user with a test set of typing characteristics of the same user or a test set of a hacker. The distance between these two sets (reference and test) should be below a certain threshold or else the user is recognized as a hacker. Neural Networks process first builds a prediction model from historical data, and then uses this model to predict the outcome of a new trial (or to classify a new observation). Although the studies tend to vary in approach from what keystroke information they utilise to the pattern classification techniques they employ, all have attempted to solve the problem of providing a robust and inexpensive authentication mechanism. Table 1 illustrates a summary of the main research approaches performed till date.

TABLE I. APPROACHES IN KEYSTROKE ANALYSIS

| Study | Classification Technique | | Users | FAR (%) | FRR (%) |
|---|---|---|---|---|---|
| Joyce & Gupta (1990) [16] | Static | Statistical | 33 | 0.25 | 16.36 |
| Leggett et al. (1991) [18] | Dynamic | Statistical | 36 | 12.8 | 11.1 |
| Brown & Rogers (1993) [6] | Static | Neural Network | 25 | 0 | 12.0 |
| Bleha & Obaidat (1993) [27] | Static | Neural Network | 24 | 8 | 9 |
| Napier et al (1995) [23] | Dynamic | Statistical | 24 | 3.8 (Combined) | |
| Obaidat & Sadoun (1997) [19] | Static | Statistical | 15 | 0.7 | 1.9 |
| | | Neural Network | | 0 | 0 |
| Monrose & Rubin (1999) [22] | Static | Statistical | 63 | 7.9 (Combined) | |
| Cho et al. (2000) [7] | Static | Neural Network | 21 | 0 | 1 |
| Ord & Furnell (2000) [25] | Static | Neural Network | 14 | 9.9 | 30 |
| Bergadano et al. (2002) [5] | Static | Statistical | 154 | 0.01 | 4 |
| Guven & Sogukpinar(2003) [13] | Static | Statistical | 12 | 1 | 10.7 |
| Sogukpinar & Yalcin(2004) [28] | Static | Statistical | 40 | 0.6 | 60 |
| Dowland & Furnell (2004) [9] | Dynamic | Neural Network | 35 | 4.9 | 0 |
| Yu & Cho (2004) [10] | Static | Neural Network | 21 | 0 | 3.69 |
| Gunetti & Picardi (2005) [12] | Static | Neural Network | 205 | 0.005 | 5 |
| Clarke & Furnell (2007) [8] | Static | Neural Network | 32 | 5 (Equal Error Rate) | |
| Lee and Cho (2007) [14] | Static | Neural Network | 21 | 0.43 (Average Integrated Errors) | |
| Pin shen The et al (2008) [27] | Static | Statistical | 50 | 6.36 (Equal Error Rate) | |

## VI. SECURITY OF KEYSTROKE DYNAMICS

So far, very little research has been conducted to analyze keystroke dynamics concerning security [4]. The application of keystroke dynamics to computer access security is relatively new and not widely used in practice. Reports on real cases of breaking keystroke dynamics authentication system do not exist. Keystroke dynamics schemes are analyzed regarding traditional attack techniques in the following section. The traditional attacks can be classified as: Shoulder Surfing, Spyware, Social Engineering, Guessing, Brute Force and Dictionary Attack

*Shoulder Surfing* A simple way to obtain a user's password is to watch them during authentication. This is called shoulder surfing. No matter if keystroke dynamics are used in the verification or identification mode, shoulder surfing is no threat for the authentication system. Password is not used in the identification case and therefore the password cannot be stolen. Only the keystroke pattern is important and decisive. In case of verification, an attacker may be able to obtain the password by shoulder surfing. However, keystroke dynamics for verification is a two-factor authentication mechanism. The keystroke pattern still has to match with the stored profile.

*Spyware* Spyware is software that records information about users, usually without their knowledge. Spyware is probably the best and easiest way to crack keystroke dynamic-based authentication systems. If a user unintentionally installs a Trojan which records all of the user's typing, keystroke latencies and keystroke durations an attacker can use this information to reproduce the user's keystroke pattern. A program could simulate the user's typing and get access to the system from the keystroke pattern. Much more research in the area is expected.

*Social Engineering* Social engineering is the practice of obtaining confidential information by manipulation of legitimate users. A social engineer will commonly use the telephone or Internet to trick people into revealing sensitive information or getting them to do something that are against typical policies. Using this method, social engineers exploit the natural tendency of a person to trust his or her word, rather than exploiting computer security holes. Phishing is social engineering via e-mail or other electronic means. On first sight, social engineering is not possible with keystroke dynamics. In the identification case there is no password that can be given away, not even on purpose. Asking for the password on the phone and pretending to be the authorized user, is not feasible. Nevertheless, phishing, social engineering via Internet, may be a way of tricking a user to give away his keystroke pattern. The attacker might portrait as a trustworthy person, asking the user to log-on to a primed website. When the user logs-on to the website the attacker might record the keystroke rhythm of the users. However, the success rate would probably be very low. The user must type his username and password several times in order to have a meaningful keystroke pattern.

*Guessing* People use common words for their passwords. The way of typing of a different user can hardly be simulated. There are just too many varieties of ways of typing on the keyboard. Guessing of typing rhymes is impossible in keystroke dynamics.

*Brute Force* In a brute force attack, an intruder tries all possible combinations of cracking a password. The more complex a password is, the more secure it is against brute force attacks. The main defense against brute force search is to have a sufficiently large password space. The password space of keystroke dynamic authentication schemes is quite large. It is nearly impossible to carry out a brute force attack against keystroke dynamics. The attack programs need to automatically generate keystroke patterns and imitate human input. If keystroke dynamics are used in a two-factor authentication mechanism, that is password and keystroke, it is almost impossible to overpower the security system.

*Dictionary Attack* A dictionary attack [4] is a technique for defeating authentication mechanism by trying to determine its pass phrase by searching a large number of possibilities. In contrast to a brute force attack, where all possibilities are searched through exhaustively, a dictionary attack only tries possibilities that are most likely to succeed, typically derived from a list of words in a dictionary. As with brute force searches, it is impractical to carry out dictionary attacks against keystroke dynamic authentication mechanisms. It is possible to use a dictionary attack which consists of general keystroke patterns, but an automated dictionary attack will be much more complex than a text based dictionary attack. Again the attack programs need to automatically generate keystroke patterns and imitate human input. Overall keystroke dynamics are less vulnerable to brute force and dictionary attacks than text based passwords.

## VII. CHALLENGES

Keystroke dynamics is a behavioral pattern exhibited by an individual while typing on a keyboard [21]. User authentication through keystroke dynamics is appealing for many reasons such as: (i) it is not intrusive, and (ii) it is relatively inexpensive to implement, since the only hardware required is the computer [12]. Unlike other physiological biometrics such as fingerprints, retinas, and facial features, all of which remain fairly consistent over long periods of time, typing patterns can be rather erratic. Even though any biometric can change over time, typing patterns have smaller time scale for changes. Not only the typing patterns is inconsistent when compared to other biometrics, a person's hands can also get tired or sweaty after prolonged periods of typing. This often results in major pattern differences over the course of a day. Another substantial problem is that typing patterns vary based on the type of the keyboard being used, the keyboard layout (i.e. qwerty or dvorak), whether the individual is sitting or standing, the person's posture if sitting, etc. The fact is that the distributed nature of keyboard biometrics also means that additional inconsistencies may be introduced into typing pattern data.

## VIII. CONCLUSION

The future of biometric technologies is promising. Biometric devices and applications continue to grow worldwide. There are several factors that will push the growth of biometric technologies. A major inhibitor of the growth of biometrics has been the cost to implement them. Moreover, increased accuracy rates will play a big part in the acceptance of biometric technologies. The development and research into biometric error testing false reject (false non-match) and false accept (false match), has been of keen interest to biometric developers. Keyboard Dynamics, being one of the cheapest forms of biometric, has great scope. In this paper an effort has been taken to give the existing approaches, security and challenges in keystroke dynamics in order to motivate the researches to further come with more novel ideas.

AUTHORS PROFILE

**Shanmugapriya. D** received the B.Sc. and M.Sc. degrees in Computer Science from Avinashilingam University for Women, Coimbatore in 1999 and 2001 respectively. And, she received the M.Phil degree in Computer Science from Manonmaniam Sundaranar University, Thirunelveli in 2003 and pursuing her PhD at Avinashilingam University for Women. She is currently working as a Lecturer in Information Technology in the same University and has eight years of teaching experience. Her research interests are Biometrics, Network Security and System Security.

**Dr. Padmavathi Ganapathi** is the Professor and Head of Department of Computer Science, Avinashilingam University for Women, Coimbatore. She has 21 years of teaching experience and one year Industrial experience. Her areas of interest include Network security and Cryptography and real time communication. She has more than 50 publications at national and International level. She is a life member of many professional organizations like CSI, ISTE, AACE, WSEAS, ISCA, and UWA.